\documentclass[aps,prd,onecolumn,amssymb]{revtex4}
\usepackage{graphicx,bm}
\usepackage[english]{babel}
\usepackage{amsmath}
\usepackage{amsfonts}
\def\beq{\begin{equation}}
\def\earn{\nonumber \end{eqnarray}}
\def\eeq{\end{equation}}

\def\Half{{\frac{1}{2}}}
\def\half{{\frac{1}{2}}}

\newcommand{\be}{\begin{equation}}
\newcommand{\ee}{\end{equation}}
\def\bear{\begin{eqnarray}}
\def\ear{\end{eqnarray}}

\begin{document}

\title{\bf {New exact cosmologies on the brane}}

\author{Artyom V. Astashenok$^{1}$ and Artyom V. Yurov$^{1}$, \\
Sergey V. Chervon$^{1,2}$ and Evgeniy V. Shabanov$^{2}$,\\
M. Sami$^{3}$}
\medskip
\affiliation{$^1$I. Kant Baltic Federal University, Kaliningrad 236041, Russia\\
$^2$Ilya Ulyanov State Pedagogical University, Ulyanovsk 432700, Russia \\
$^3$Centre of Theoretical Physics, Jamia Millia Islamia, New Delhi 110025, India}

\begin{abstract}
{We develop a method for constructing exact cosmological solutions in
brane world cosmology. New classes of cosmological solutions on Randall -- Sandrum brane are obtained. The superpotential and Hubble parameter are represented in quadratures. These solutions have inflationary phases under general assumptions and also describe an exit from the inflationary phase without a fine tuning of the parameters. Another class solutions can describe the current phase of accelerated expansion with or without possible exit from it.} %
\end{abstract}

\maketitle

\section{Introduction}

Brane world scenario have been proposed more then decade ago and this scenario attracted many attention. The reason of such attention includes the hope to describe observed accelerated expansion of the Universe.

To our knowledge the work by Binetrruy, Deffaet and Langlois \cite{bidela9905} was the first work where brane cosmology, different from standard Friedmann cosmology, have been proposed. It is interesting to mention that in the first work on brane cosmology at once it was much attention to searching of exact solutions. This situation is a diametrically opposite to that in cosmological inflation theory where about a decade approximated solution have been under study till the works \cite{barrow90,muslimov90}. Further development of the brane cosmology scenarios was performed in works \cite{bdel9910,mukohyama9911, vollick9911,kraus9910,ida9912} where few exact solutions have been found. In in the work \cite{mutesh9912} it was shown the relation between exact solutions in \cite{kraus9910,ida9912} and \cite{bdel9910,mukohyama9911}. Namely it was given the explicit coordinate transformation which proved the equivalence between this two solutions,i.e. both solutions
represent the same spacetime in different coordinate systems.

In consideration of inflation scenario in brane cosmology it is important to analyze scalar field solutions with self-interacting potential. There are number of works devoted to this issue. We will mention few of them where investigation of exact solutions was carried out. In the article \cite{pauls02} it was found exact solution at high energy limit when $H^2$ proportional to $\rho^2$; canonical and tachyon fields were analyzed there. A general thick brane with a scalar field was analysed in the work \cite{bromei04}, where restrictions on the scalar potential was obtained and exact solutions for stepwise potentials of different shapes was found. A general scalar field and barotropic fluid during the early stage of
a brane-world where the Friedmann constraint is dominated by the square of the energy density was studied in the work \cite{mizuno-etal04}. In this article we present the method of exact solution construction and new classes of exact solutions obtained by superpotential method (see \cite{yurov-etal11} and literature cited therein). Fist application of this method for Randal -- Sandrum brane cosmology was perfomed in the work \cite{chesam09}.

The article is organized as follow. In section II we present the model and method of exact solution construction for it. Section III devoted to solutions obtained for given superpotential. In section IV we present new exact solutions based on given evolution of the scalar field. In section V we discuss viable cosmological models obtained in the article.

\section{Cosmological models on the FRW brane: methods of solutions construction}

As an alternative to the FRW cosmology let us consider the simplest brane model in which spacetime is homogeneous and isotropic along three spatial dimensions, being our 4-dimensional universe an infinitesimally thin wall, with constant spatial curvature, embedded in a 5-dimensional spacetime (\cite{Sahni},\cite{Langlois}). In the Gaussian normal coordinate system, for the brane which is located at $y=0$, one gets
\be
ds^{2}=-n^{2}dt^{2}+a^{2}(t,y)\gamma_{ij}dx^{i}dx^{j}+dy^{2},
\ee
where $\gamma_{ij}$ is the maximally 3-dimensional metric.
Let $t$ be the proper time on the brane ($y=0$), then $n(t,0)=1$. Therefore, one gets the FRW metric on the brane
\be
ds^{2}_{|y=0}=-dt^{2}+a^{2}(t,0)\gamma_{ij}dx^{i}dx^{j}.
\ee
The 5-dimensional Einstein equations have the form
\be
R_{AB}-\frac{1}{2}g_{AB}R=\chi^{2}T_{AB}+\Lambda_{4} g_{AB},
\ee
where $\Lambda_{4}$ is the bulk cosmological constant, $\chi^2=8\pi G^{(5)}/c^{4}$, $G^{(5)}$ is the gravitational constant in 5-dimensional spacetime. The next step is to write the total energy momentum tensor $T_{AB}$ on the brane as
\be
T^{A}_{B}=S^{A}_{B}\delta(y),
\ee
with $S^{A}_{B}=\mbox{diag}(-\rho_{b},p_{b},p_{b},p_{b},0)$, where $\rho_{b}$ and $p_{b}$ are the total brane energy density and pressure, respectively.

One can now calculate the components of the 5-dimensional Einstein tensor which solve Einstein's equations. One of the crucial issues here is to use appropriate junction conditions near $y=0$. These reduce to the following two relations:
\be
\frac{dn}{ndy}_{|y=0+}=\frac{\chi^{2}}{3}\rho_{b}+\frac{\chi^{2}}{2}p_{b},\qquad \frac{da}{ady}_{|y=0+}=-\frac{\chi^{2}}{6}\rho_{b}.
\ee
After some calculations, one  obtains the following result
\be
H^{2}=\chi^{4}\frac{\rho_{b}^{2}}{36}+\frac{\Lambda_{4}}{6}-\frac{k}{a^{2}}+\frac{\mathcal{E}}{a^{4}}.
\ee
This expression is valid on the brane  only. Here $H=\dot{a}(t,0)/a(t,0)$ and $\mathcal{E}$ is an
arbitrary integration constant. The energy conservation equation is correct, too,
\be
\dot{\rho_{b}}+3\frac{\dot{a}}{a}(\rho_{b}+p_{b})=0.
\ee
Now, let $\rho_{b}=\rho+\lambda_{b}$, where $\lambda_{b}$ is the brane tension. Further we consider the fine-tuned brane with $\Lambda_{4}=\lambda_{b}^{2}\chi^{4}/6$ and the case of flat spacetime ($k=0$):
\be \label{BREQ-1}
\frac{\dot{a}^{2}}{a^{2}}=\frac{\lambda\chi^{4}}{6}\frac{\rho}{3}
\left(1+\frac{\rho}{2\lambda_{b}}\right)+\frac{\mathcal{E}}{a^{4}}.
\ee
In what follows we will consider a single brane model which mimics GR at present but differs from it at late times.
We set $M_{p}^{-2}=8\pi G=\sigma\chi^4/6$. For simplicity, we set $\mathcal{E}=0$ (the term with $\mathcal{E}$ is usually called ``dark radiation''). In fact, setting $\mathcal{E}\neq 0$ does not lead
to additional solutions on a radically new basis, in the framework of our approach. Eq.~(\ref{BREQ-1}) can be simplified to
\be \label{frw-w-br}
\frac{\dot{a}^{2}}{a^{2}}=\frac{\rho}{3M_{p}^{2}}\left(1+\frac{\rho}{2\lambda_{b}}\right).
\ee
One can see that Eq.~(\ref{frw-w-br}), for $\rho<<|\lambda_{b}|$, differs insignificantly from the FRW equation.  The brane model with a positive tension has been discussed in \cite{Liddle},\cite{Sami},\cite{Sami-2} in the context of  the unification of early- and late-time acceleration eras.
The braneworld model with a negative tension and a time-like extra dimension can be regarded as being dual to the Randall-Sundrum model (\cite{Sahni-2},\cite{Randall},\cite{Copeland}). Note that, for this model, the Big Bang singularity is absent. And this fact does not depend upon whether or not matter violates the energy conditions (\cite{Asht}). This same scenario has also been used to construct cyclic models for the universe \cite{Cyclic}.

One can assume that in our epoch the $\rho/2\lambda<<1$ and so there is no significant difference between the brane model and FRW cosmology. But the universe evolution in the future or in past, for brane cosmology, can in fact differ from such convenient cosmology, due to the non-linear dependence of the expansion rate on the energy density.

One can reduce the field equation to the slow-roll form
\beq\label{f}
3HU=-W'_\phi,
\eeq
with substitution
\beq\label{f1}
W=V+\Half U^2,~~U(\phi)=\dot\phi
\eeq
Then Friedmann equation (\ref{frw-w-br}) one can write down in terms of the superpotential $W$
\beq\label{e1}
H^2=\frac{1}{3M_p^2}W\left(1+\frac{W}{2\lambda_b}\right).
\eeq

Considering $H$ as positive and inserting (\ref{e1}) into (\ref{f}) one can obtain

\beq\label{w-u}
\frac{\sqrt{3}}{M_p}U(\phi)\sqrt{W\left(1+\frac{W}{2\lambda_b}\right)}=
-W'_\phi .
\eeq

This is the key equation for further progress. For given $W(\phi)$ as function of scalar field one can define the dependence of scalar field from time inversing the following relation:
\be\label{tphi}
t-t_{0}=-\frac{\sqrt{3}}{M_p}\int  \frac{d\phi}{W'_\phi}\sqrt{W\left(1+\frac{W}{2\lambda_b}\right)}
\ee

In frames of this approach the physical potential is
\be\label{Vphi}
V(\phi)=W(\phi)-\frac{M_{p}^2}{6}W'^{2}_\phi\left(W(\phi)\left(1+\frac{W(\phi)}{2\lambda_{b}}\right)\right)^{-1}
\ee

One can also define $U(\phi)$ as function of scalar field. In this case the integration (\ref{w-u}) via separation of $W$ and $\phi$  leads to the superpotential and Hubble parameter presentation in quadratures

\bear
\label{superW+}
\sqrt{W}=\sqrt{2\lambda_b} \sinh\left( -\sqrt{\frac{3}{2\lambda_b}}\frac{1}{2M_P}
\int U(\phi)d\phi \right),~~W>0, \\
\label{superW-}
\sqrt{-W}=\sqrt{2\lambda_b} \cosh\left( -\sqrt{\frac{3}{2\lambda_b}}\frac{1}{2M_P}
\int U(\phi)d\phi \right),~~W<0, \\
\label{superH}
H=\sqrt{\frac{\lambda_b}{6}}\frac{1}{M_P} \sinh
\left(-\sqrt{\frac{3}{2\lambda_b}}\frac{1}{M_P}\int U(\phi)d\phi \right)
\ear

Note that the argument in (\ref{superH}) should be positive. Therefore $ \int U(\phi)d\phi $ should be always negative.

The physical potential $V$ can be obtained from the superpotential definition (\ref{f1})
\beq\label{potV}
V(\phi)=2\lambda_b \sinh^2 \left( -\sqrt{\frac{3}{2\lambda_b}}\frac{1}{2M_P}
\int U(\phi)d\phi \right)
 -\half U(\phi)^2.
\eeq

Thus to obtain the examples of exact solutions one can suggest the functional dependence a scalar field $\phi$ on cosmic time $t$ and evaluate the integral $\int U(\phi)d\phi $:
$$
\int U(\phi) d\phi=\int U^2({t}) dt.
$$

Standard calculation leads to the following formula for $N(\phi)$:
\beq
\textsc{N}(\phi)=-\sqrt{\frac{\lambda_b}{2}}M^{-2}_P \int \sqrt{W\left(1+\frac{W}{2\lambda_b}\right)}\sqrt{\left(\frac{W}{\lambda_b}+
1\right)^2+1}\frac{d\phi}{W'}
\eeq

To understand the period with accelerating Universe expansion let us calculate $ \frac{\ddot{a}}{a}=\dot{H}+H^2$. For given $W(\phi)$ we have the following simple relation for acceleration parameter:
\be\label{acceler}
\frac{\ddot{a}}{a}=-\frac{1}{6}\frac{W'^{2}_{\phi}}{W}+\frac{W}{3M^{2}_{P}}+
\frac{W}{6\lambda_{b}}\left(\frac{W}{M^{2}_{P}}-\frac{1}{2}\frac{W'^{2}_{\phi}}{W(1+W/2\lambda_{b})}\right)
\ee
The first two terms corresponds to the case of usual FRW cosmology. The last appears due to the brane tension. Combining  (\ref{acceler}) with
(\ref{tphi}) one can analyze the behavior $\ddot{a}/{a}$ as function of time.

If we define the function $U(\phi)$ it is convenient to use the result
\beq
\label{accel}
\frac{\ddot{a}}{a}=-\frac{U^2}{2M_P^2}\cosh \left(-\sqrt{\frac{3}{2\lambda_b}}\frac{1}{M_P}\int U(\phi)d\phi \right)
+\frac{\lambda_b}{6M_P^2}\sinh^2 \left(-\sqrt{\frac{3}{2\lambda_b}}\frac{1}{M_P}\int U(\phi)d\phi \right)
\eeq
To investigate changing of the acceleration sign let us represent (\ref{accel}) in the following form
\beq\label{eq-z}
\frac{6M_P^2}{\lambda_b}\frac{\ddot{a}}{a}=Z^2 - \frac{3U^2}{\lambda_b}Z - 1,~~
Z=\cosh \left(-\sqrt{\frac{3}{2\lambda_b}}\frac{1}{M_P}\int U(\phi)d\phi \right)
\eeq
Taking into account that $Z>1$ we omit the the root
\beq
Z_2 =\frac{3U^2}{2\lambda_b}-\sqrt{\frac{9U^4}{4\lambda^2_b}+1}
\eeq
as it is less then unity. For the next root
\beq\label{z1}
Z_1=\frac{3U^2}{2\lambda_b}+\sqrt{\frac{9U^4}{4\lambda^2_b}+1}
\eeq
it is easy to check that $Z_1>1$ for any value of $U^2$. Therefore we can imply that there is only one point during the evolution when deceleration have been changed to acceleration. So in the framework of scalar field cosmology on the brane we have only one inflationary period. If it is early inflation then once again as in FRW cosmology we need to solve the problem of exit from inflation (for example using generation of the particle after inflation ends). On the other hand, if it is later inflation, we may suggest existence one more additional specie at the early stage of Universe evolution which will be responsible for early inflation.

\section{Solutions with given $W(\phi)$}

Let's consider two simple superpotentials. The first case is
\be
W=\frac{m^2\phi^2}{2}
\ee
One can easy derive the dependence $t(\phi)$. The result is
$$
t-t_{0}=-\frac{\sqrt{6\lambda_b}}{2m^{2}}M_P^{-1}\left[ \mbox{arcsinh} \left( \frac{m \phi}{2\sqrt{\lambda_b}}\right)
+\frac{m\phi}{2\sqrt{\lambda_b}}\sqrt{1+\frac{m^2\phi^2}{4\lambda_b}} \right]
$$
For simplicity we put $\phi(t_{0})=0$. At $t\rightarrow\infty$ scalar field $\phi\rightarrow-\infty$. One can consider the moment $t<t_{0}$ for this moment $\phi>0$. The universe acceleration is
\be
\frac{\ddot{a}}{a}=-\frac{m^{2}}{12}+\frac{m^{2}\phi^{2}}{6M_{P}^{2}}+\frac{m^{2}\phi^{2}}{12\lambda_{b}}\left(\frac{m^{2}\phi^{2}}{2M_{P}^{2}}-
\frac{1}{2}\frac{m^{2}}{1+m^{2}\phi^{2}/4\lambda_{b}}\right)
\ee
For scale factor as function of scalar field we have
\be
a=a_{0}\exp\left(\pm\frac{\phi^2}{8M_{P}^{2}}\left(2+\frac{m^2\phi^2}{4\lambda_{b}}\right)\right)
\ee
One can choose the sign $-$ for $t<0$ (i.e. $\phi>0$) and $+$ for $t>0$ ($\phi<0$). The moment $t=-\infty$ corresponds to Big Bang singularity then universe expands so that $\ddot{a}>0$ then non-inflationary phase follows. Then universe again expands with acceleration. The asymptotic of solution is \be
a(t)\sim a_{0}\exp\left(\frac{\lambda_{b}m^2}{2M_{P}^{2}}(t-t_{0})^{2}\right).
\ee
The potential of scalar field can be obtained from (\ref{Vphi}):
\be
V(\phi)=\frac{m^2\phi^2}{2}-\frac{M_{P}^{2}m^2}{3}\left(1+\frac{m^2\phi^2}{4\lambda_{b}}\right)^{-1}.
\ee
At $t\rightarrow\infty$ this potential corresponds to free scalar field with mass $m$.

For case of $\phi-4$ superpotential
\be
W(\phi)=\frac{\lambda\phi^4}{4}
\ee
we have the following link between time and scalar field:
\be
t-t_{0}=-\frac{\sqrt{3}}{2M_{P}\sqrt{\lambda}}\left(\cosh\eta-\cosh\eta_{0}+\ln\tanh\frac{\eta}{2}-\ln\tanh\frac{\eta_{0}}{2}\right),
\ee
$$
\quad \eta=\mbox{arcsinh}\left(\left(\frac{\lambda}{8\lambda_{b}}\right)^{1/2}\phi^2\right).
$$
The scalar field at the moment $t=t_{0}$ is $\phi_{0}=\left(\frac{8\lambda_{b}}{\lambda}\right)^{1/4}\sinh^{1/2}\eta_{0}$ where $\eta_{0}$ is arbitrary constant. At $t\rightarrow\infty$ the scalar field $\phi\rightarrow 0$. From expression for universe acceleration
\be
\frac{\ddot{a}}{a}=-\frac{2\lambda}{3}\phi^2+\frac{\lambda}{12M_{P}^{2}}\phi^4+\frac{\lambda^2\phi^6}{24\lambda_{b}}\left(\frac{1}{4M_{P}^{2}}\phi^2-
\frac{2}{1+\lambda\phi^4/8\lambda_{b}}\right)
\ee
one can see that exit from inflation occurs. The scale factor as function of scalar field
\be
a=a_{0}\exp\left(\pm\frac{\phi^2}{8M_{P}^{2}}\left(1+\frac{\lambda\phi^4}{24\lambda_{b}}\right)\right)
\ee
The negative sign corresponds to universe starting from $a=0$ ($\phi=\infty$ at $t=-\infty$) and expanding with acceleration prior to some moment of time when exit from inflation occurs.

\section{Solutions with given $\phi(t)$}

To obtain exact solutions we can use the dynamics of the scalar
field considered in many works. Let us start from the simplest
ones.

\subsection{Logarithmic evolution of the scalar field} 

Let
$$
\phi = A \ln (\lambda t).
$$

The solutions are
\bear
\label{superW+ln}
\sqrt{W}=\sqrt{2\lambda_b} \sinh\left( -\sqrt{\frac{3}{2\lambda_b}}\frac{1}{2M_P}
\left(C_1
-\frac{A^2}{t}\right)\right),~~W>0, \\
\label{superW-ln}
\sqrt{-W}=\sqrt{2\lambda_b} \cosh \left(-\sqrt{\frac{3}{2\lambda_b}}\frac{1}{2M_P}
\left(C_1-\frac{A^2}{t}\right)\right),~~W<0, \\
\label{superH_ln}
H=\sqrt{\frac{\lambda_b}{6}}\frac{1}{M_P} \sinh
\left(-\sqrt{\frac{3}{2\lambda_b}}\frac{1}{M_P}\left(C_1
-\frac{A^2}{t}\right)\right).
\ear

The superpotential's and physical potential's presentation in terms of scalar field can
be obtained using given dependance $ \phi = A \ln (\lambda t).$

\bear
\label{superW+ln-phi}
\sqrt{W}=\sqrt{2\lambda_b} \sinh\left( -\sqrt{\frac{3}{2\lambda_b}}\frac{1}{2M_P}
\left(C_1
-A^2\lambda\exp(-\phi/A)\right)\right),~~W>0, \\
\label{superW-ln-phi}
\sqrt{-W}=\sqrt{2\lambda_b} \cosh \left(-\sqrt{\frac{3}{2\lambda_b}}\frac{1}{2M_P}
\left(C_1-A^2\lambda\exp(-\phi/A)\right)\right),~~W<0, \\
\label{V_5.1}
V(\phi)=2\lambda_b \sinh^2 \left( -\sqrt{\frac{3}{2\lambda_b}}\frac{1}{2M_P}
\left(C_1-A^2\lambda\exp(-\phi/A)\right)\right)
 -\frac{A^2 \lambda^2}{2}e^{-2\phi/A}.
\ear

The potential $V(\phi)$ is depicted on Fig. 1. Thus we have obtained the exact formulas for given evolution of the scalar field. We know that
for solutions under consideration we may have only one inflection point for scalar factor $a$ associated with the equation (\ref{z1}).
Using the definition for $Z$ (\ref{eq-z}) one can obtain the equation for time corresponding to inflection point:

\begin{equation}
\label{intersection_5.1}
\frac{3A^2}{2\lambda_b t^2}+\sqrt{\frac{9A^4}{4\lambda^2_b t^4}+1}=
\cosh \left(\sqrt{\frac{3}{2\lambda_b}}\frac{A^2}{M_P t} \right)
 \end{equation}

\begin{figure}
    \begin{center}
        \includegraphics[scale=0.75]{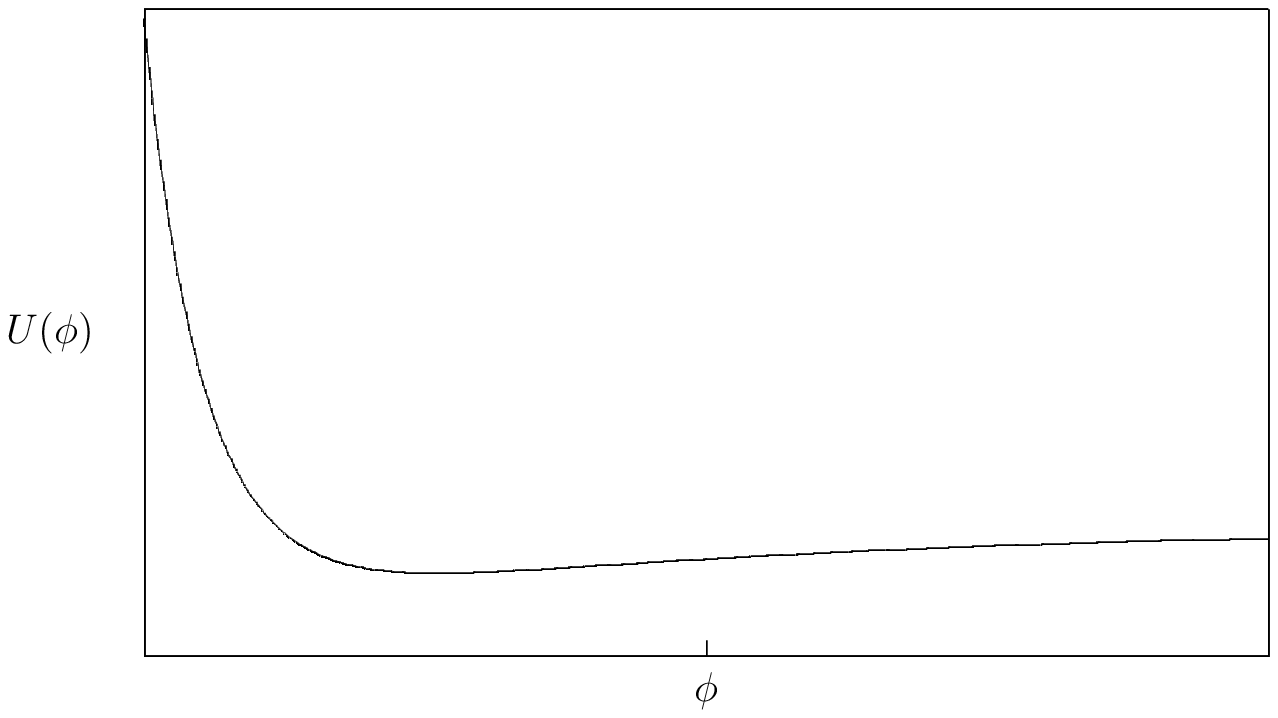}
    \end{center}
\label{fig.1}
\caption{The potential of scalar field for logarithmic evolution of scalar field.}
\end{figure}

It is easy to see that this equation will be true when $t \rightarrow \infty $.
We can analyze this equation numerically to find the finite time $ t< \infty $ which will correspond to deceleration changes to acceleration.
One can state that this time is about $t_{i}\approx0.286{\lambda_{b}^{-1/2}}$ in the units with $M_P=1$ and for $A=1$. Therefore we can use this approximate result to set this time equal to beginning of (early or later) inflation. From the other hand we can analyze the transition from brane cosmology to Friedmann one by tending the brane tension to infinity $\lambda_b \rightarrow \infty $. The results are:

\bear
\nonumber\sqrt{\lambda_b} \rightarrow \infty: \\
\sqrt{W}=-\sqrt{3}\frac{1}{2M_p}\left(C_1
-\frac{A^2}{t}\right)=-\sqrt{3}\frac{1}{2M_p}\left(C_1
-A^2\lambda e^{-\phi/A}\right), \\
\sqrt{-W} = \infty, \\
H=-{\frac{1}{\sqrt{6}M_p}}\left(\sqrt{\frac{3}{2}}\frac{1}{M_p}\left(C_1 -\frac{A^2}{t}\right)\right), \\
V(\phi)=\infty.
\ear

Here we can define the scale factor with power law -- exponential behavior
\beq
a(t)=\exp \left(-\frac{C_1 t}{2M_P^2}\right)t^{\frac{2A^2\lambda^2}{2M_P^2}}.
\eeq

We carefully analyzed the solution in this section. To simplify presentation of the next solutions let us represent equations (\ref{superW+})-(\ref{potV}) in the following way
\bear
\label{superW+m}
\sqrt{W}=\sqrt{2\lambda_b} \sinh\left( -\sqrt{\frac{3}{2\lambda_b}}\frac{1}{2M_P}
\left[ F(\phi)+C_1\right]\right),~~W>0, \\
\label{superW-m}
\sqrt{-W}=\sqrt{2\lambda_b} \cosh\left( -\sqrt{\frac{3}{2\lambda_b}}\frac{1}{2M_P}
 \left[F(\phi)+C_1 \right]\right),~~W<0, \\
\label{superH-m}
H=\sqrt{\frac{\lambda_b}{6}}\frac{1}{M_P} \sinh
\left(-\sqrt{\frac{3}{2\lambda_b}}\frac{1}{M_P}\left[ F(\phi(t))+C_1\right] \right)\\
\label{potV-m}
V(\phi)=2\lambda_b \sinh^2 \left( -\sqrt{\frac{3}{2\lambda_b}}\frac{1}{2M_P}
\left[ F(\phi)+C_1\right]\right)
 -\half [U(\phi)]^2.
\ear

Using the general formulas (\ref{superW+m})-(\ref{potV-m}) we will display new solutions by putting values for $F(\phi)$ and $U(\phi)$. Also we will use general formulas for transition to Friedmann Universe (by setting $\lambda_b \rightarrow \infty $ ) below
\bear
\nonumber\sqrt{\lambda_b} \rightarrow \infty: \\
\sqrt{W}=-\frac{\sqrt{3}}{2M_p}\left(F(\phi)+C_1 \right), \\
\sqrt{-W} = \infty, \\
H=-{\frac{1}{2M_p^2}}\left(F(\phi (t))+C_1 \right), \\
V(\phi)=-\frac{3}{4M_P}(F(\phi)+C_1)^2-  \frac{1}{2}[U(\phi)]^2
\ear

\subsection{Power law evolution}

Let
$$
\phi =A t^s,~~s\neq 0, ~~s\neq 1/2.
$$

The solutions are represented by functions
\beq
\label{U5.2}
F(\phi)=\frac{A^{-1/s}s^{2}}{2s-1}\phi^{2-1/s}=\frac{A^2s^2}{2s-1}t^{2s-1},~~U^2(\phi)= A^2s^2 \left[\frac{\phi}{A}\right]^{\frac{2s-2}{s}}.
\eeq

The equation for the time corresponding to inflection point is
\beq
\label{intersection_5.2}
\frac{3A^2s^2}{2\lambda_b}t^{2s-2}+\sqrt{\frac{9A^4s^4}{4\lambda_b}t^{4s-4}+1}=\cosh \left( -\sqrt{\frac{3}{2\lambda_b}}\frac{A^2s^2}{(2s-1)M_P}t^{2s-1}\right)
\eeq

By tending the brane tension to infinity $\lambda_b \rightarrow \infty $ we obtain

\bear
\nonumber\sqrt{\lambda_b} \rightarrow \infty:\\
\sqrt{W}=-\frac{\sqrt{3}}{2M_p}\left(\frac{A^2s^2}{2s-1}
\left[\frac{\phi}{A}\right]^{\frac{2s-1}{s}}+C_1\right),\\
\sqrt{-W}=\infty,\\
H=-{\frac{1}{\sqrt{6}M_p}}\left(\sqrt{\frac{3}{2}}\frac{1}{M_p}\left(C_1
+\frac{A^2s^2}{2s-1}
t^{\frac{2s-1}{s}}\right)\right),\\
V(\phi)=\infty.
\ear

In the case when $s=1/2$ we have

\beq
F(\phi)=\frac{A^2}{2}\ln \frac{\phi}{A},~~U^2(\phi)= \frac{A^4}{4\phi^2}
\eeq

The potential of scalar field in this case is depicted on Fig. 2.

The equation for the time corresponding to inflection point is
\beq
\label{intersection_5.2_s=1/2}
\frac{3A^2}{8t\lambda_b}+\sqrt{\frac{9A^4}{32t^2\lambda_b^2}+1}=\frac{1}{2}\left(t^{-B}+t^B\right),~~
B=\sqrt{\frac{3}{2\lambda_b}}\frac{A^2}{4M_P}
\eeq

This time is about $t\approx4.11\lambda_{b}^{-1/2}$ in units $M_{P}=A=1$. In this moment the deceleration begins.

\begin{figure}
    \begin{center}
        \includegraphics[scale=0.75]{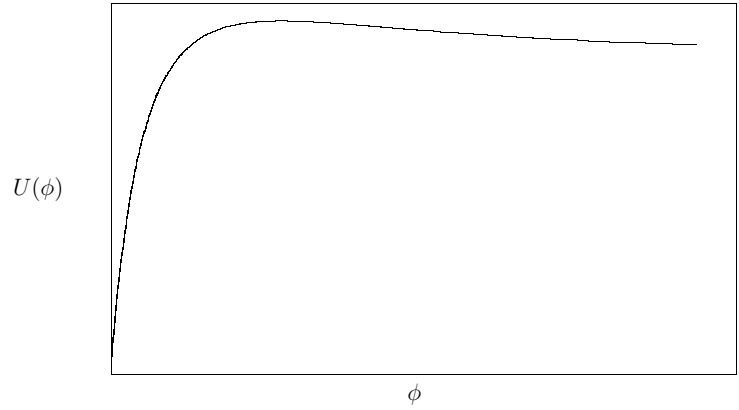}
    \end{center}
\label{fig.3}
\caption{The potential of scalar field for $\phi=At^{1/2}$.}

\end{figure}

\subsection{Exponential evolution}

Let

$$
\phi = A e^{-\lambda t}.
$$

The solutions represented by functions
\beq
\label{U5.3}
F(\phi)=-\frac{\phi^2\lambda}{2}=-\frac{A^2\lambda}{2}e^{-2\lambda t},~~U^2(\phi)= \lambda^2\phi^2
\eeq

The potential of scalar field is presented on Fig. 3.

\begin{figure}
    \begin{center}
        \includegraphics[scale=0.75]{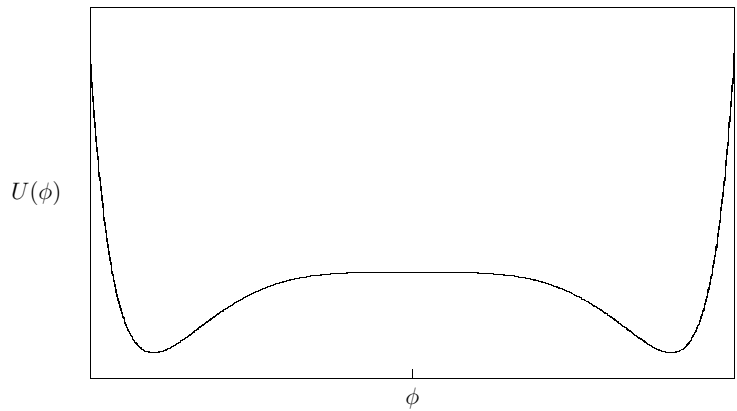}
    \end{center}
\label{fig.6}
\caption{The potential of scalar field for $\phi=A\exp(-\lambda t)$.}

\end{figure}

The equation for the time corresponding to inflection point is
\beq
\label{intersection_5.3}
\frac{3A^2\lambda^2}{2\lambda_b}e^{-2\lambda t}+\sqrt{\frac{9A^4\lambda^4}{4\lambda_b^2}e^{-4\lambda t}+1}=\cosh \left(\sqrt{\frac{3}{2\lambda_b}}\frac{A^2\lambda}{M_P}e^{-2\lambda t}\right),~~
\eeq
For inflection time we have two different solutions:
i) when $\lambda > 0 $ $\ddot{a} \rightarrow 0$ at $t \rightarrow \infty$ (and $\ddot{a}>0$ at $0<t<\infty$); ii) when $\lambda=-\sqrt{\lambda_{b}}<0 $ $\ddot{a} = 0$ at $t\approx 0.879\lambda_{b}^{-1/2}$ ($M_{p}=A=1$). At this time acceleration changes to deceleration i.e. we have the inflation phase during finite time.

By tending the brane tension to infinity $\lambda_b \rightarrow \infty $ we obtain
\bear
\sqrt{\lambda_b} \rightarrow \infty:\\
\sqrt{W}=-\sqrt{3}\frac{1}{2M_p}\left(C_1
-\frac{A^2 \lambda}{2}e^{-2\lambda t}\right),\\
\sqrt{-W}=\infty,\\
H=-{\frac{1}{\sqrt{6}M_p}}\left(\sqrt{\frac{3}{2}}\frac{1}{M_p}\left(C_1
-\frac{A^2 \lambda}{2}e^{-2\lambda t}\right)\right),\\
V(\phi)=\infty.
\ear

The solutions above have been obtained earlier in \cite{chesam09} with a slightly different form. Here we presented them in more suitable way and gave detailed analysis. The solutions below are obtained first time and based on the scalar field evolutions considering in cosmology \cite{barrow94,che04}. Investigation of cosmological parameters for such evolution of scalar field was performed in the work \cite{chefom08}.

\subsection{New classes of solutions}

\subsubsection{$\phi=A\ln(\tanh(\lambda t))$}

The solution is represented by functions
\bear
F(\phi)=A^2\lambda\left(2\cosh(\phi/A)\right)=A^2\lambda\left(\tanh(\lambda t)+\coth(\lambda t)\right),\\
\label{U5.4.1}
U^2(\phi)= \frac{A^2\lambda^2(1-\exp(\frac{2\phi}{A}))^2}{\exp(\frac{2\phi}{A})}
\ear

The potential of scalar field is depicted on Fig. 4.

\begin{figure}
    \begin{center}
        \includegraphics[scale=0.75]{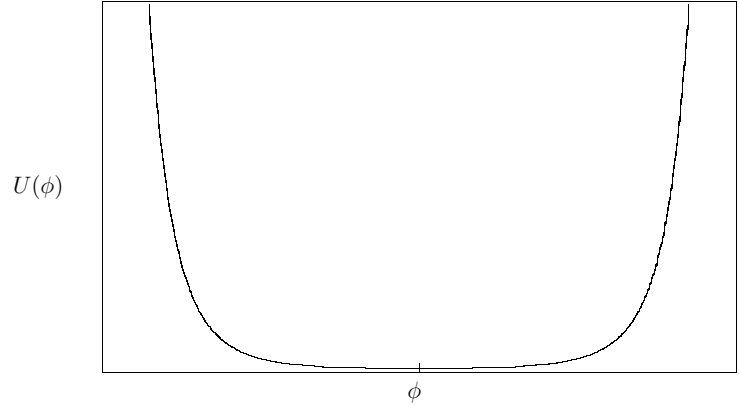}
    \end{center}
\label{fig.8}
\caption{The scalar field potential for $\phi=A\ln(\tanh(\lambda t))$}

\end{figure}

The inflection point for the scale factor $a(t)$ can be obtained as a solution of the following equation:

\bear
\label{intersection_5.4.1}
\frac{3A^2\lambda^2}{2\lambda_b\cosh^2(\lambda t)\sinh^2(\lambda t)}+\sqrt{\frac{9A^4\lambda^4}{4\lambda^2_b\cosh^4(\lambda t)\sinh^4(\lambda t)}+1}= \nonumber\\
=\cosh\left(\sqrt{\frac{3}{2\lambda_b}}\frac{A^2\lambda}{M_p}\left(\tanh(\lambda t)+\coth(\lambda t)\right)\right)
\ear
This equation has solutions at $\lambda<0.82\lambda_{b}^{1/2}$ (for $A=M_{P}=1$). The accelerated expansion begins at $t=t_{1}$ and ends at $t=t_{2}$. In table 1 the duration of inflation stage are given for various $\lambda$.

\begin{table}
\begin{tabular}{|c|c|}
\hline
  $\lambda/\lambda^{1/2}$ &  \\
  \hline
  0.8 & $0.38<t\lambda_{b}^{1/2}<0.53$ \\
  0.6 & $0.315<t\lambda_{b}^{1/2}<0.382$  \\
  0.4 & $0.295<t\lambda_{b}^{1/2}<1.2$ \\
  0.2 & $0.286<t\lambda_{b}^{1/2}<2.31$ \\
  \hline
\end{tabular}
\caption{The duration of accelerated expansion for model $\phi=A\ln(\tanh(\lambda t))$ for various $\lambda$.}
\end{table}

\subsubsection{$\phi=A\ln(\tan(\lambda t))$}

The solution is represented by formulas

\bear
F(\phi)=A^2\lambda(2\sinh(\phi/A))=A^2\lambda(\tan(\lambda t)-\cot(\lambda t)),\\
\label{U5.4.2}
U^2(\phi)= \frac{A^2\lambda^2(1+\exp(\frac{2\phi}{A}))^2}{\exp(\frac{2\phi}{A})}
\ear

On Fig. 5 one can see the dependence of potential from scalar field.

\begin{figure}
    \begin{center}
        \includegraphics[scale=0.75]{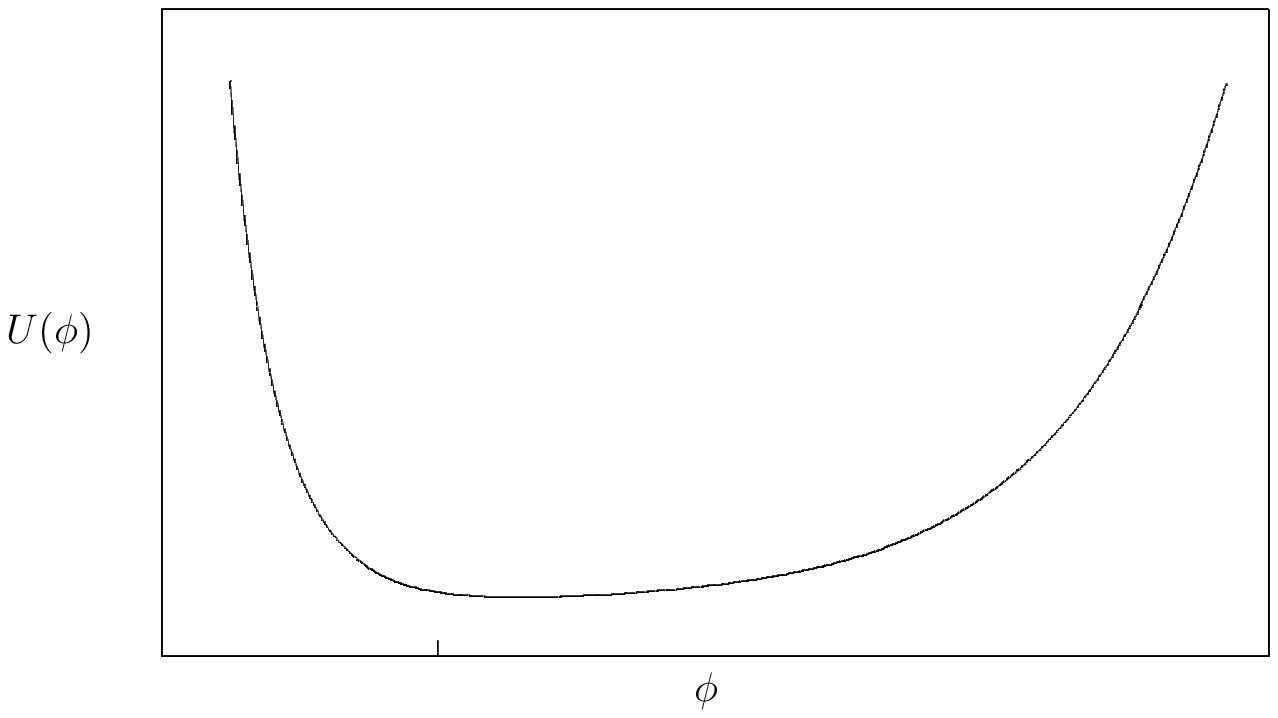}
    \end{center}
\label{fig.10}
\caption{The scalar field potential for $\phi=A\ln(\tan(\lambda t))$}

\end{figure}

The time for inflection point can be found as a solution of equation:
\bear
\label{intersection_5.4.2}
\frac{3A^2\lambda^2}{2\lambda_b\cos^2(\lambda t)\sin^2(\lambda t)}+\sqrt{\frac{9A^4\lambda^4}{4\lambda^2_b\cos^4(\lambda t)\sin^4(\lambda t)}+1}= \nonumber\\
=\cosh\left(\sqrt{\frac{3}{2\lambda_b}}\frac{A^2\lambda}{M_p}\left(\tan(\lambda t)-\cot(\lambda t)\right)\right)
\ear

We have the same situation as in previous case. For various values of $\lambda$ we have two roots corresponding to moment of beginning of inflation and to moment of exit from inflation. In table 2 these results are given for various values of $\lambda$.

\begin{table}
\begin{tabular}{|c|c|}
\hline
  $\lambda/\lambda^{1/2}$ &  \\
  \hline
  0.6 & $0.265<t\lambda_{b}^{1/2}<2.355$ \\
  0.8 & $0.251<t\lambda_{b}^{1/2}<1.712$ \\
  1.0 & $0.239<t\lambda_{b}^{1/2}<1.335$ \\
  1.2 & $0.225<t\lambda_{b}^{1/2}<1.083$  \\
  1.4 & $0.215<t\lambda_{b}^{1/2}<0.907$ \\
  \hline
\end{tabular}
\caption{The duration of accelerated expansion for model $\phi=A\ln(\tan(\lambda t))$ for various $\lambda$.}
\end{table}

\subsubsection{$\phi=A/\sinh(\lambda t)$}

The solution is described by the functions

\bear
F(\phi)=A^2\lambda\coth(\lambda t)\frac{1}{3}\biggl[\frac{\phi^2}{A^2}-1\biggr]=A^2\lambda\coth(\lambda t)\frac{1}{3}\biggl[\frac{1}{\sinh^2(\lambda t)}-1\biggr],\\
\label{U5.4.3}
U^2(\phi)= \frac{\lambda^2}{A^{2}}\phi^4\left(1+\left(\frac{A}{\phi}\right)^{2}\right)
\ear

The potential of scalar field is depicted on Fig. 6.

\begin{figure}
    \begin{center}
        \includegraphics[scale=0.75]{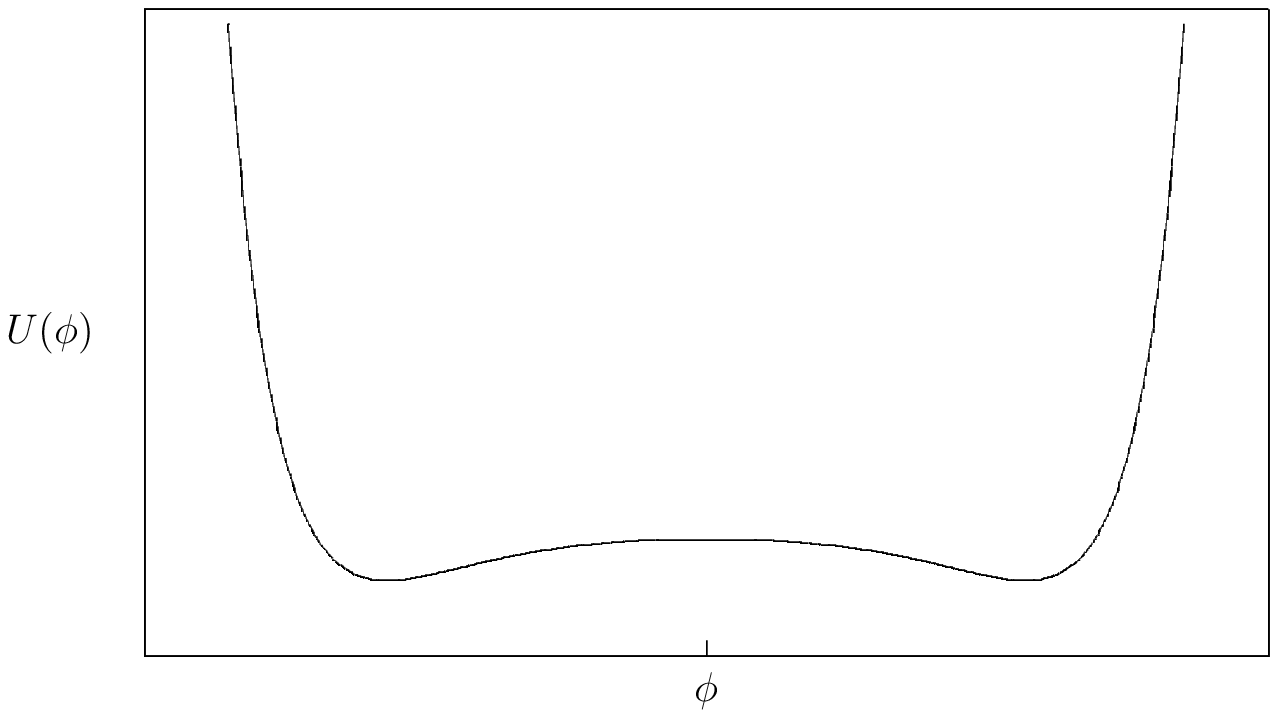}
    \end{center}
\label{fig.12}
\caption{The scalar field potential for $\phi=A/\sinh(\lambda t)$}

\end{figure}

The inflection point can be found from corresponding equation
\bear
\label{intersection_5.4.3}
\frac{3A^2\lambda^2\cosh^2(\lambda t)}{2\lambda_b\sinh^4(\lambda t)}+\sqrt{\frac{9A^4\lambda^4\cosh^4(\lambda t)}{4\lambda^2_b\sinh^8(\lambda t)}+1}= \nonumber\\
=\cosh\left(\sqrt{\frac{3}{2\lambda_b}}\frac{A^2\lambda\cosh(\lambda t)}{3M_p}\biggl[\frac{1}{\sinh^2(\lambda t)}-1\biggr]\right)
\ear

Once again for various values of $\lambda$ we have phase of accelerated expansion during finite time. In table 3 one can see results for various values of $\lambda$.

\begin{table}
\begin{tabular}{|c|c|}
\hline
  $\lambda/\lambda^{1/2}$ &  \\
  \hline
  0.6 & $0.665<t\lambda_{b}^{1/2}<3.5$ \\
  0.8 & $0.545<t\lambda_{b}^{1/2}<2.63$ \\
  1.0 & $0.468<t\lambda_{b}^{1/2}<2.11$ \\
  1.2 & $0.414<t\lambda_{b}^{1/2}<1.765$  \\
  1.4 & $0.374<t\lambda_{b}^{1/2}<1.52$ \\
  \hline
\end{tabular}
\caption{The duration of accelerated expansion for model $\phi=A/\sinh(\lambda t)$ for various $\lambda$ ($A=M_{P}=1$).}
\end{table}

\subsubsection{$\phi=A\arctan(\exp(\lambda t))$}

The solution is represented by functions

\bear
F(\phi)=-\frac{A^2\lambda\cos^2(\phi/A)}{2}=-\frac{A^2\lambda}{2(1+\exp(2\lambda t)},\\
\label{U5.4.4}
U^2(\phi)= \frac{A^2\lambda^2\tan^2(\frac{\phi}{A})}{(1+\tan^2(\frac{\phi}{A}))}.
\ear

The potential of scalar field is presented on Fig. 7.

\begin{figure}
    \begin{center}
        \includegraphics[scale=0.75]{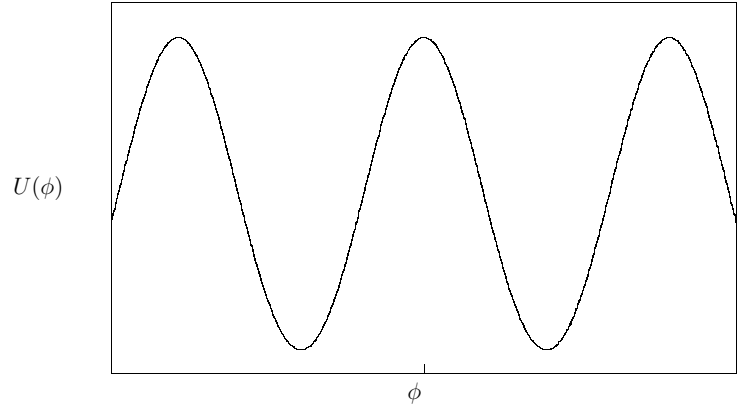}
    \end{center}
\label{fig.14}
\caption{The scalar field potential for $\phi=A\arctan(\exp(\lambda t))$}

\end{figure}

From the equation

\bear
\label{intersection_5.4.4}
\frac{3A^2\lambda^2\exp(2\lambda t)}{2\lambda_b(1+\exp(2\lambda t))^2}+\sqrt{\frac{9A^4\lambda^4\exp(4\lambda t)}{4\lambda^2_b(1+\exp(2\lambda t))^4}+1}= \nonumber\\
=\cosh\left(\sqrt{\frac{3}{2\lambda_b}}\frac{A^2\lambda}{2M_p(1+\exp(2\lambda t))}\right)
\ear

one can find that for $\lambda > 0$ $\ddot{a}>0$ ($0<t<\infty$) and for $\lambda < 0$ we have the moment when deceleration begins. For example for $\lambda=-\lambda_{b}^{1/2}$ we have $t\approx 0.093\lambda_{b}^{-1/2}$.

\subsubsection{$\phi=A\sin^{-1}(\lambda t)$}

The solution is described by formulas

\bear
F(\phi)=A^2\lambda\cot(\lambda t)\frac{1}{3}\biggl[1-\frac{1}{\sin^2(\lambda t)}\biggr]=A^2\lambda\cot(\lambda t)\frac{1}{3}\biggl[1-\frac{1}{\sin^2(\lambda t)}\biggr],\\
\label{U5.4.5}
U^2(\phi)=\frac{\lambda^2}{A^{2}}\phi^4\left(1-\left(\frac{A}{\phi}\right)^{2}\right).
\ear

On Fig. 8 one can see the potential of scalar field.

\begin{figure}
    \begin{center}
        \includegraphics[scale=0.75]{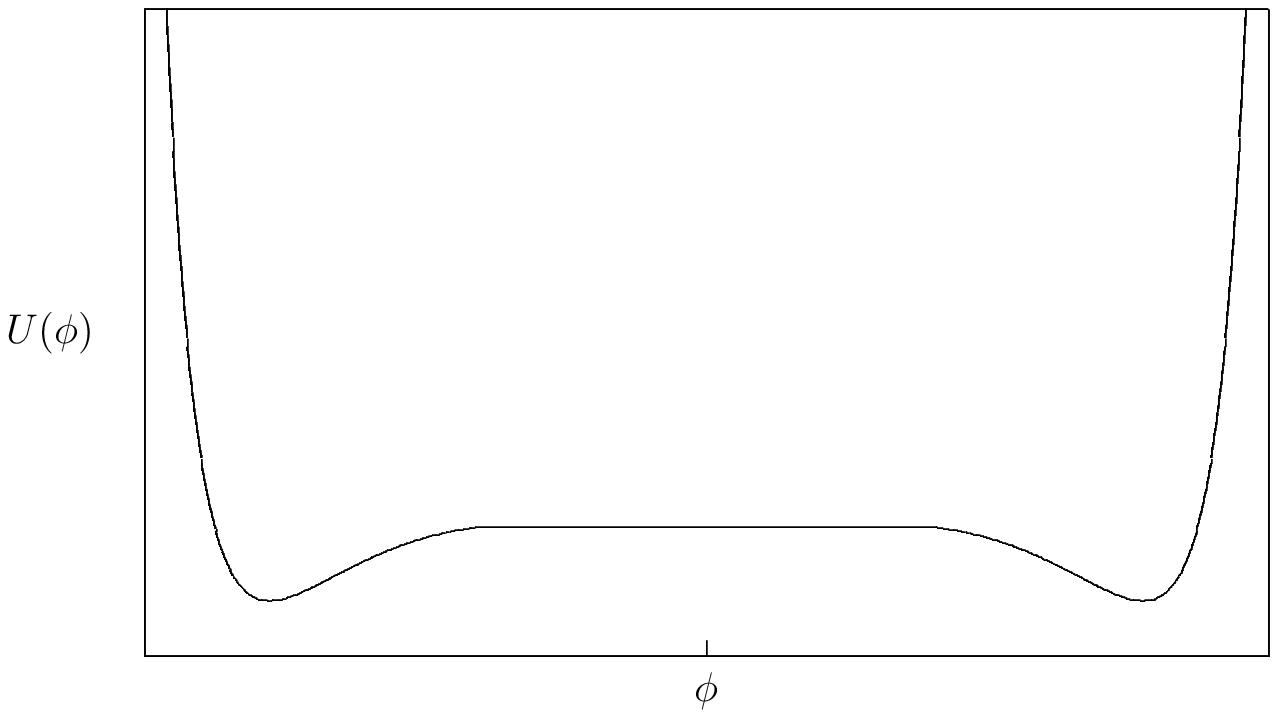}
    \end{center}
\label{fig.16}
\caption{The scalar field potential for $\phi=A\sin^{-1}(\lambda t)$}

\end{figure}

The inflection point can be found from equation

\bear
\label{intersection_5.4.5}
\frac{3A^2\lambda^2\cos^2(\lambda t)}{2\lambda_b\sin^4(\lambda t)}+\sqrt{\frac{9A^4\lambda^4\cos^4(\lambda t)}{4\lambda^2_b\sin^8(\lambda t)}+1}= \nonumber\\
=\cosh\left(\sqrt{\frac{3}{2\lambda_b}}\frac{A^2\lambda\cot(\lambda t)}{3M_p}\biggl[1-\frac{1}{\sin^2(\lambda t)}\biggr]\right).
\ear

For $\lambda=\lambda_{b}^{1/2}$ we have the phase of accelerated expansion for $t>0.4\lambda_{b}^{-1/2}$.

\section{Discussions}

In this paper we have discussed a simple method of construction of exact solutions for cosmological equations on RS brane. Despite simplicity, the method allows for acquirement of solutions characterized by interesting properties. These solutions have inflationary phases under quite general assumptions. This is an indication that inflationary regime seems a common occurrence in cosmology not requiring any special initial conditions. We have the following cases:

(i) accelerated expansion begins at some moment of time and lasts forever (logarithmic evolution of scalar field, $\phi \thicksim \sin^{-1}(\lambda t)$). In principle current observed acceleration can be described by this case.

(ii) acceleration take place before $t<t_{f}$ when deceleration starts (power and exponential evolution of scalar field or $\phi\thicksim \arctan(\exp(\lambda t))$). These solutions may correspond to inflation in beginning of cosmological evolution and exit from inflation phase.

(iii) acceleration occurs during interval $t_{i}<t<t_{f}$ ($\phi\thicksim \ln(\tanh(\lambda t))$, $\phi\thicksim \ln(\tan(\lambda t))$, $\phi\thicksim 1/\sinh(\lambda t)$). The interpretation of such solutions is obvious: we have the possible exit from inflation or current accelerated expansion or initial inflation rolling to later slow inflation (with further exit from it).

Therefore these models can not only describe inflation but also describe an exit from the inflationary phase without a fine tuning of the parameters. This fact can be seen as evidence of the existence of a realistic model that contains an inflationary phase in the early universe stage and a current phase of accelerated expansion.

\end{document}